\documentclass[12pt, a4paper]{article}
\usepackage{amsmath,amssymb,amsfonts}
\usepackage[colorlinks]{hyperref}
\usepackage{graphicx}

\begin{document}

\title{Properties of the ground state of electronic excitations in carbon-like nanocones}

\author{Yurii A. Sitenko$^1$ , Volodymyr M. Gorkavenko$^2$\\
\it \small ${}^{1}$Bogolyubov Institute for Theoretical Physics,
  National Academy of Sciences of Ukraine,\\ \it \small
  14-b Metrologichna str., Kyiv 03143,
 Ukraine\\
 \it \small ${}^{2}$Department of Physics, Taras Shevchenko
National University of Kyiv,\\ \it \small 64 Volodymyrs'ka str.,
Kyiv
 01601, Ukraine}

\date{}
\maketitle

\abstract{On the basis of the
continuum model for long-wavelength charge carriers, originating in the tight-binding approximation for the nearest-neighbour interaction of atoms in the crystalline lattice, we consider
quantum ground-state effects of electronic excitations in Dirac materials with two-dimensional
monolayer honeycomb structures warped into nanocones by a disclination; the nonzero size of the disclination is taken into account, and a boundary condition at the edge of the disclination is chosen to ensure self-adjointness of the Dirac-Weyl Hamiltonian operator. We show that the quantum ground-state effects are independent of the disclination size and find circumstances when they are independent of a parameter of the boundary condition. The magnetic flux circulating in the angular direction around the nanocone apex and the pseudomagnetic flux directed orthogonally to the nanocone surface are shoun to be induced in the ground state.}

\phantom{hvhv}


\section{Introduction}

A wealth of new phenomena in micro- and
nanophysics, suggesting possible
applications to technology and industry, is promised by a synthesis
in this century of strictly two-dimensional atomic crystals
(for instance, a monolayer of carbon atoms, graphene, \cite{Nov,Ge}). The electronic states near the Fermi level in these crystals are characterized by the linear and isotropic dispersion relation, with the density of states at the Fermi level being strictly zero. Condensed matter systems with such a behavior of electronic excitations are known as the two-dimensional Dirac materials comprising a diverse set ranging from honeycomb crystalline structures (graphene \cite{Nov}, silicene and germanene \cite{Cah}, phosphorene \cite{Liu}) to high-temperature $d$-wave superconductors, superfluid phases of helium-3
and topological insulators, see review in \cite{Weh}. Using the tight-binding approximation for the nearest-neighbour interaction in the crystalline lattice, an effective long-wavelength description of electronic excitations can be given in terms of a continuum model which is based on the Dirac-Weyl equation for massless electrons in $2+1$-dimensional space-time, with the role of velocity of light $c$ played by Fermi velocity $v\approx c/300$ \cite{Di,Sem}.

Freely suspended samples of crystalline monolayers are not exactly plane surfaces, but possesss ripples which are due to the appearance of topological defects in a crystallne lattice: disclinations and
 disclination dipoles (dislocations). A single disclination warps a sheet of the crystallne lattice, giving it the shape of a cone. The squared length element of the conical surface is
\begin{equation}\label{1.2}
    ds^2= dr^2+\nu^{-2} r^2 d\varphi^2, \quad 0\leq \varphi < 2\pi,
\end{equation}
where $\nu=(1-\eta)^{-1}$, and $2\pi\eta$ is the deficit angle. Conical spaces (i.e. $3$-dimensional spaces with a $2$-dimensional section given by \eqref{1.2}) emerge in a field rather different from condensed matter physics -- in cosmology. The early universe in the process of its cosmological expansion is likely to undergo a series of phase transitions with spontaneous breakdown of continuous symmetries, and a vortex-like topological defect which is formed in the aftermath of such a transition is known under the name of a cosmic string, see reviews in \cite{Vi, Hi}. Starting with a random tangle, the cosmic string network evolves into two distinct sets: the stable one which consists of several long, approximatly straight strings spanning the horizon volume and the unstable one which consists of a variety of string loops decaying by gravitational radiation. A straight infinitely long cosmic string in its rest frame is characterized by an outer space with the transverse section given by \eqref{1.2}. Parameter
$\eta$ is related to the mass per unit length of the cosmic string, hence it is positive, and the present-day astrophysical
observations restrict its values to range
$0< \eta < 10^{-6}$ (see, e.g., \cite{Bat}).

On the contrary, in the case of conically-shaped crystalline monolayers, parameter
$\eta$ takes both positive and negative discrete values of order 1 and even larger: a disclination obtained by deleting atoms from the crystalline lattice results in the positive deficit angle, whereas a disclination obtained by adding atoms into the crystalline lattice results in the negative deficit (i.e. proficit) angle. For instance, in the case of the honeycomb lattice of graphene, silicene, germanene or phosphorene, a natural way of
producing the apex of a nanocone is by substitung some of hexagons by
pentagons (positive deficit angle) or heptagons (negative deficit angle); thus,
$\eta=N_d/6$, where $N_d$ is an integer which is smaller than 6. A general disclination in
the honeycomb lattice is obtained by substituting a
hexagon by a polygon with $6-N_d$ sides; polygons with $N_d>0$
($N_d<0$) induce locally positive (negative) curvature at the apex, whereas the crystalline sheet is locally flat away from the disclination, as is the conical surface away from the apex. In the case of nanocones with
$N_d>0$, the value of $N_d$ is related to apex angle $\delta$,
$\sin\frac\delta2=1-\frac{N_d}{6}$, and $N_d$ counts the number of
sectors of the value of $\pi/3$ which are removed from the crystalline
sheet. If $N_d<0$, then $-N_d$ counts the number of such sectors
which are inserted into the crystalline sheet. Certainly, polygonal
defects with $N_d>1$ and $N_d<-1$ are mathematical abstractions,
as are cones with a pointlike apex. In reality, the defects are
smoothed, and $N_d>0$ counts the number of the pentagonal defects
which are tightly clustered producing a conical shape; carbon
nanocones with the apex angles
$\delta=112.9^\circ,\,83.6^\circ,\,60.0^\circ,\,38.9^\circ,\,19.2^\circ$,
which correspond to the values $N_d=1,\,2,\,3,\,4,\,5$, were
observed experimentally, see \cite{Heiberg} and references therein.
Theory also predicts an infinite series of the saddle-like nanocones
with quantity $-N_d$ counting the number of the heptagonal defects
which are tightly clustered forming the saddle centre. Saddle-like nanocones
serve as an element which is necessary for joining parts of carbon
nanotubes of differing radii.

Another distinction from the case of cosmic strings is in the intertwinement of valleys, as well as sublattices, in the case of disclinations corresponding to odd values of $N_d$. It seems reasonable to identify a matrix exchanging both the sublattice and valley indices with
$\gamma^5$. Hence, the relevant bundle connection corresponding to the gauge axial vector field appears, describing the pseudomagnetic vortex with flux related to the deficit angle. This is in contrast to the case of cosmic strings, where the relevant bundle connection correspond to the gauge vector field describing the vortex with flux unrelated to the deficit angle.

In the present paper, we consider the quantum ground-state effects of electronic
excitations in honeycomb crystalline monolayer structures with disclinations corresponding to $N_d = \pm 1, \, \pm 2, \pm 3, \, 4, \, 5, -6$. A crucial point is a choice of the boundary condition at the location of the disclination. The previous consideration \cite{SiV7,SiV1, SiV2} was neglecting the transverse size of the disclination, treating it as a pointlike one. We are now tackling the problem more carefully by taking the finite size of the disclination into account, imposing the most general boundary condition at the disclination edge, and then going to  the physically sensible limit of the nanocone size exceeding considerably the disclnation size. This more physical approach allows us to specify the boundary condition with more definiteness. We find out that the pseudomagnetc field directed orthogonally to the nanocone surface is induced in the ground state, whereas the electric charge is not; the magnetic field circulating in the angular direction around the nanocone apex is induced in the ground state in cases $N_d = \pm 2, \, -6$ only.

\section{Continuum model description of electronic excitations in monolayer atomic crystals with a disclination}

Electronic excitations in a plane sheet of the honeycomb crystalline lattice are described in terms of a four-component wave function,
\begin{equation}\label{10}
\psi=\left(\psi_A^{(I)},\,\psi_B^{(I)},\,\psi_A^{(II)},\,\psi_B^{(II)}\right)^T,
\end{equation}
where subscripts $A$ and $B$ correspond to two sublattices and superscripts $(I)$ and $(II)$ correspond to two valleys (inequivalent Fermi points). As was noted in Introduction, in the framework of the long-wavelength continuum model, the wave function of electronic excitations  satisfies the Dirac--Weyl equation,
\begin{equation}\label{9}
\left({\rm i}\partial_0 - H \right)\psi = 0, \quad  H=-{\rm i}\hbar v \left(\alpha^1\partial_1+\alpha^2\partial_2 \right).
\end{equation}
The generating elements of the Clifford algebra of anticommuting matrices in $3+1$-dimensional space-time can be chosen as
\begin{equation}\label{12}
    \gamma^0=\tau^0\sigma^3,\quad
    \alpha^1=-\tau^0\sigma^2,\quad
\alpha^2=\tau^3\sigma^1,\quad
\alpha^3=\tau^1\sigma^1,
\end{equation}
where $\sigma^0$ and $\sigma^{j}$ ($\tau^0$ and $\tau^{j}$) are the unity and Pauli matrices with the sublattice (valley) indices, and $j=1,2,3$. Defining
$\gamma^5=-{\rm i}\alpha^1\alpha^2\alpha^3$, one gets
\begin{equation}\label{13}
\gamma^5=-\tau^2\sigma^2.
\end{equation}
A rotation by angle $\vartheta$ in the plane of a honeycomb lattice sheet is
implemented by operator $\exp(i\vartheta\Sigma)$, where
\begin{equation}\label{14}
    \Sigma=\frac{1}{2{\rm i}}\alpha^1\alpha^2=\frac12 \tau^3\sigma^3
\end{equation}
is the pseudospin playing here the role of the operator of spin component which is orthogonal to the plane. The honeycomb lattice is invariant under a rotation by
$2\pi$, but is not invariant under a rotation by $\pi$. The parity transformation can be introduced as a  rotation by $\pi$, which is simultaneously supplemented by the exchange of both the sublattice and valley indices \cite{Gus},
\begin{equation}\label{15}
P\psi=\left(\psi_B^{(II)},\,\psi_A^{(II)},\,\psi_B^{(I)},\,\psi_A^{(I)}\right)^T
\end{equation}
with
\begin{equation}\label{15a}
P =2\Sigma R, \quad [P,H]_+=[R,H]_-=0;
\end{equation}
in representation \eqref{12} we obtain
\begin{equation}\label{14a}
P =\alpha^3, \quad R=\gamma^5.
\end{equation}
The wave function is chosen as a section of a bundle with spin connection
$-2\Sigma$, i.e. it obeys condition
\begin{equation}\label{27}
    \psi(\varphi+2\pi)=-\psi(\varphi).
\end{equation}

If a defect with $N_d= \pm1$ is inserted at the origin, then condition  \eqref{27} is changed to the
M\"{o}bius-strip-type condition:
\begin{equation}\label{28}
  \psi(\varphi+2\pi) = \pm{\rm i} R \psi(\varphi), \quad \psi(\varphi+4\pi)=-\psi(\varphi).
\end{equation}
For a general defect with $N_d < 6$, the condition takes form
\begin{equation}\label{30}
    \psi(\varphi+2\pi)=-\exp\left(-{\rm i}\frac\pi2 N_d R\right)\psi(\varphi),
\end{equation}
while the Hamitonian operator for electronic excitations in a conical surface with the squared length element given by \eqref{1.2} takes form
\begin{equation}\label{1.14}
H=-{\rm i}\hbar v\left[\alpha^r\left(\partial_r+\frac{1}{2r}\right)+\alpha^\varphi\partial_\varphi \right],
\end{equation}
where
\begin{equation}\label{1.14a}
\alpha^r=\alpha_r=-\tau^0\sigma^{2}, \quad
\alpha^\varphi=\frac{\nu}r\tau^3\sigma^{1}, \quad \alpha_\varphi=\frac{r^2}{\nu^2}\,\alpha^\varphi.
\end{equation}
By performing a singular gauge transformation, we arrive at the wave function obeying condition \eqref{27} and the Hamiltonian operator involving bundle connection $\Phi(2\pi \hbar v)^{-1}$
\cite{SiV7}:
\begin{equation}\label{1.14b}
H=-{\rm i}\hbar v\left[\alpha^r\left(\partial_r+\frac{1}{2r}\right)+\alpha^\varphi\left(\partial_\varphi-{\rm i}\frac{{\Phi}}{2\pi \hbar v}\right)\right],
\end{equation}
where
\begin{equation}\label{1.10a}
\Phi= 3\pi \hbar v(1- \nu^{-1}) R, \quad
\nu=(1-N_d/6)^{-1};
\end{equation}
note that in the case of cosmic strings quantity $\Phi$ is the flux of a gauge vector field corresponding to the generator of a spontaneously broken continuous symmetry.

Next, by performing in addition a unitary transformation, we arrive at the representation with both $R$ and $P$ diagonal,
\begin{equation}\label{1.15}
\gamma^5=R=\tau^3\sigma^0,
\quad
\alpha^3=P=\tau^0\sigma^{3},
\quad
\gamma^0=\tau^1\sigma^1,
\end{equation}
while relations \eqref{14} and \eqref{1.14a} are maintained. The initial representation with diagonal $\gamma^0$, see \eqref{12}, can be denoted as the standard one, and it has been chosen to be diagonal in both the sublattice and the valley indices, see \eqref{10}. The final representation with diagonal $\gamma^5$, see \eqref{1.15}, can be denoted as the chiral one, and it mixes up sublattices, as well as valleys.

Using the chiral representation, we decompose the solution to the stationary
Dirac-Weyl equation, $ H \psi_{E}({\bf x})=E \psi_{E}({\bf x})$, with $H$ given by \eqref{1.14b} and \eqref{1.10a} as
\begin{equation}\label{1.16}
\psi_E(\textbf{x}) = \sum_{n \in \mathbb{Z}}
                   \left(\begin{array}{c}
                   f_{n,+}(r,E )e^{ {\rm i} (n+1/2)\varphi} \\
                   g_{n,+}(r,E )e^{ {\rm i} (n+1/2)\varphi} \\
                   f_{n,-}(r,E )e^{ {\rm i} (n-1/2)\varphi} \\
                   g_{n,-}(r,E )e^{ {\rm i} (n-1/2)\varphi}
                    \end{array}\right),
\end{equation}
where the radial functions satisfy the system of first-order differential equations
\begin{equation}\label{1.17}
 \left\{
 \begin{array}{c}
\hbar v\left[-\partial_r +\frac1r (\pm \nu n-\nu+1)\right] f_{n, \pm}(r,E) =E g_{n, \pm}(r,E) \\
\hbar v\left[\partial_r +\frac1r (\pm \nu n-\nu+2)\right]
g_{n, \pm}(r,E) =E f_{n, \pm}(r,E)
\end{array}
\right\};
\end{equation}
thus a component of definite chirality, $+$ or $-$, is a superposition of components with definite sublattice and valley indices.

Quantum effects in the ground state of electronic excitations
comprise the induced electric charge density:
\begin{equation}\label{1.7aa}
q(\textbf{x})=-\frac{e}2 \int\limits_{-\infty}^\infty \frac{dE\,
E}{\hbar^2 v^2} \mbox{$\psi$}^\dag _E(\textbf{x}) \mbox{$\psi$}
_E(\textbf{x}),
\end{equation}
the induced electric current density:
\begin{equation}\label{1.7ad}
\textbf{j}(\textbf{x})=-\frac{e}{2} \int\limits_{-\infty}^\infty
\frac{dE\, E}{\hbar^2 v} \mbox{$\psi$}^\dag _E(\textbf{x})
\mbox{\boldmath $\alpha$}   \mbox{$\psi$}_E(\textbf{x});
\end{equation}
the induced parity-breaking condensate density:
\begin{equation}\label{1.7ab}
\rho(\textbf{x})=-\frac{1}2 \int\limits_{-\infty}^\infty \frac{dE\,
E}{\hbar^2 v^2} \mbox{$\psi$}^\dag _E(\textbf{x}) P\, \mbox{$\psi$}
_E(\textbf{x}),
\end{equation}
and the induced $R$-current density:
\begin{equation}\label{1.7ac}
\textbf{j}^{R}(\textbf{x})=-\frac12 \int\limits_{-\infty}^\infty
\frac{dE\, E}{\hbar^2 v} \mbox{$\psi$}^\dag _E(\textbf{x})
\mbox{\boldmath $\alpha$} R\,  \mbox{$\psi$}_E(\textbf{x}).
\end{equation}
The magnetic field strength, $\textbf{B}(\textbf{x})$, is also induced in the ground state, as a consequence of
the Maxwell equation,
\begin{equation}\label{1.8}
\mbox{\boldmath $\partial$}\times \textbf{B}(\textbf{x}) =
\frac1v\, \textbf{j}(\textbf{x}),
\end{equation}
as well as does the pseudomagnetic field strength, $\textbf{B}^R(\textbf{x})$, which is a consequence of
the analogue of the Maxwell equation,
\begin{equation}\label{1.9}
\mbox{\boldmath $\partial$}\times \textbf{B}^R (\textbf{x}) =
\frac1v\, \textbf{j}^{R}(\textbf{x});
\end{equation}
the use of term ``pseudomagnetic'' is justified because $R$ coincides with $\gamma^5$; due to this, also the $R$-current can be regarded as an axial current.

Using \eqref{1.14a}, \eqref{1.15} and \eqref{1.16}, one gets $j^{R}_r =0$ immediately and, with more careful analysis (see the beginning of Section 4), $j^{R}_3 =0$, where
\begin{equation}\label{1.1a}
j^{R}_3(r) = - \frac 12 \int\limits_{-\infty}^\infty \frac{dE\,
E}{\hbar^2 v} \sum_{n \in \mathbb{Z}} [f_{n,+}^2(r,E) - g_{n,+}^2(r,E) - f_{n,-}^2(r,E)+ g_{n,-}^2(r,E)].
\end{equation}
Thus, the only component of the induced ground-state $R$-current,
\begin{equation}\label{1.19}
j^{R}_\varphi(r) = - \frac r\nu \int\limits_{-\infty}^\infty \frac{dE\,
E}{\hbar^2 v} \sum_{n \in \mathbb{Z}} [f_{n,+}(r,E) g_{n,+}(r,E) +
f_{n,-}(r,E) g_{n,-}(r,E)],
\end{equation}
is independent of the angular variable. The induced ground-state pseudomagnetic
field strength is also independent of the angular variable, being directed orthogonally to the conical surface,
\begin{equation}\label{1.20}
B^{R}_3(r) =\frac{\nu}v \int\limits_r^{r_{\rm max}}
\frac{dr'}{r'} \, j^{R}_\varphi(r')+ B^{R}_3(r_{\rm max}),
\end{equation}
with total flux
\begin{equation}\label{1.21}
\Phi_{\rm I}^R = \frac{2\pi}\nu \int\limits_{r_0}^{r_{\rm max}} dr\, r B^{R}_3(r),
\end{equation}
where it is assumed without a loss of generality that a nanocone is of a rotationally invariant shape with $r_{\rm max}$
being its radius and $r_0$ being the radius of a disclination, $r_{\rm max} \gg r_0$ in the physically sensible case.

Turning to the induced ground-state electric charge and parity-breaking condensate, their densities
are also independent of the angular variable:
\begin{equation}\label{1.7aa1}
q(r)=-\frac{e}2 \int\limits_{-\infty}^\infty \frac{dE\,
E}{\hbar^2 v^2}\sum_{n \in \mathbb{Z}}  [f_{n,+}^2(r,E)
+g_{n,+}^2(r,E) + f_{n,-}^2(r,E)+ g_{n,-}^2(r,E)]
\end{equation}
and
\begin{equation}\label{1.7aa2}
\rho(r)=-\frac{1}2 \int\limits_{-\infty}^\infty \frac{dE\,
E}{\hbar^2 v^2}\sum_{n \in \mathbb{Z}}  [f_{n,+}^2(r,E)
-g_{n,+}^2(r,E) + f_{n,-}^2(r,E)- g_{n,-}^2(r,E)].
\end{equation}
Appropriately, one can define total charge
\begin{equation}\label{1.7aa3}
Q=\frac{2\pi}{\nu} \int\limits_{r_0}^{r_{max} } dr\,r\, q(r)
\end{equation}
and total $P$-condensate
\begin{equation}\label{1.7bb3}
C=\frac{2\pi}{\nu} \int\limits_{r_0}^{r_{max} } dr\,r\, \rho(r).
\end{equation}
Note that the induced ground-state condensate of pseudospin $\Sigma$ \eqref{14} is proportional to $j^{R}_3$ \eqref{1.1a} and, thus, is vanishing.

As to the induced ground-state electric current, note an evident relation, $j_r = 0$, and a less evident one (substantiated in the beginning of Section 4), $j_\varphi = 0$, where
\begin{equation}\label{1.19a}
j_\varphi(r) = - \frac {e r}{\nu} \int\limits_{-\infty}^\infty \frac{dE\,
E}{\hbar^2 v} \sum_{n \in \mathbb{Z}} [f_{n,+}(r,E) g_{n,+}(r,E) -
f_{n,-}(r,E) g_{n,-}(r,E)];
\end{equation}
hence, the only nonvanishing component is directed orthogonally to the conical surface and is related to the $P$-condensate:
\begin{equation}\label{1.19a}
j_3(r) = e v \rho(r).
\end{equation}
The total electric current,
\begin{equation}\label{1.7cc3}
J_3=\frac{2\pi}{\nu} \int\limits_{r_0}^{r_{max} } dr\,r\, j_3(r),
\end{equation}
is appropriately related to the total $P$-condensate:
\begin{equation}\label{1.19f}
J_3 = e v C.
\end{equation}
The induced ground-state magnetic
field strength is also independent of the angular variable, being directed in the conical surface along a circle with an apex in its center,
\begin{equation}\label{1.20b}
B_\varphi(r) =-\frac1{\nu v} \int\limits_r^{r_{\rm max}}
dr' \, r' \, j_{3}(r')+ B_\varphi(r_{\rm max}).
\end{equation}
Its total flux is
\begin{equation}\label{1.20c}
\Phi_{\rm I} = \int\limits_{r_0}^{r_{max} } dr\, B_\varphi(r).
\end{equation}

Concluding this Section, note that we are considering the ground-state characteristics which are diagonal in chiralities. The nondiagonal ones (for instance, the $\gamma^0$-condensate) are proportional, as follows from \eqref{1.16}, either to $\cos\varphi$ or to $\sin\varphi$ and, thus, vanish upon averaging over the angular variable.

\section{Self-adjointness and choice of boundary conditions}

Let us note first, that \eqref{1.14b} is not
enough to define the Hamiltonian operator rigorously in a
mathematical sense. To define an operator in a unambiguous way, one
has to specify its domain of definition. Let the set of functions $\psi$
be the domain of definition of operator $H$, and the set of functions
$\tilde \psi$ be the domain of definition of its adjoint, operator
$H^\dag$. Then the operator is Hermitian (or symmetric in
mathematical parlance),
\begin{equation}\label{1.10}
\int\limits_X d^2x \sqrt{g} \,{\tilde\psi}^\dag (H\psi)=
\int\limits_X d^2x \sqrt{g}\,(H^\dag \tilde \psi)^\dag \psi,
\end{equation}
if relation
\begin{equation}\label{1.11}
-{\rm i} \int\limits_{\partial X} d \mbox{\boldmath l}\,
{\tilde \psi}^\dag \mbox{\boldmath $\alpha$} \psi =0
\end{equation}
is valid; here functions $\psi(\textbf{x})$ and $\tilde
\psi(\textbf{x})$ are defined in space $X$ with boundary $\partial
X$. It is evident that condition \eqref{1.11} can be satisfied by
imposing different boundary conditions for $\psi$  and $\tilde
\psi$.  But, a nontrivial task is to find a possibility that a
boundary condition for $\tilde \psi$ is the same as that for $\psi$;
then the domain of definition of $H^\dag$ coincides with that of
$H$, and operator $H$ is self-adjoint (for a review of the Weyl-von
Neumann theory of self-adjoint operators see \cite{Neu,Ree}). The action
of a self-adjoint operator results in functions belonging to its
domain of definition only, and a multiple action and functions of
such an operator,  for instance, the resolvent and evolution operators,
can be consistently defined. Thus, in the case of a surface of radius
$r_{\rm max}$ with a deleted central disc of radius $r_0$, we have to ensure
the validity of relations
\begin{equation}\label{3.1}
\left.\tilde\psi^\dag\alpha^r \psi\right|_{r=r_0}=0, \quad \left.\tilde\psi^\dag\alpha^r \psi\right|_{r=r_{\rm max}}=0,
\end{equation}
meaning that the quantum matter excitations do not penetrate outside. It is implied that functions $\psi$
and $\tilde \psi$ are differentiable and square-integrable. As $r_{\rm max}\rightarrow \infty$, they conventionally
turn into differentiable functions corresponding to the continuum, and the condition at $r=r_{\rm max}$ yields
 no restriction at $r_{\rm max}\rightarrow \infty$, whereas the condition at $r=r_0$ yields
\begin{equation}\label{3.2}
\left. \psi\right|_{r=r_0}=\left. K\psi \right|_{r=r_0}, \quad
\left. \tilde\psi\right|_{r=r_0}=\left. K\tilde \psi
\right|_{r=r_0},
\end{equation}
where $K$ is a matrix (element of the Clifford algebra in $2+1$-dimensional space-time) which obeys
condition
\begin{equation}\label{3.3}
K^2=I
\end{equation}
and   without a loss of generality can be chosen to be Hermitian; in
addition, it has to obey either condition
\begin{equation}\label{3.4}
[K,\alpha^r]_+=0,
\end{equation}
or condition
\begin{equation}\label{3.5}
[K,\alpha^r]_-=0.
\end{equation}
One can simply go through four linearly independent elements of the
Clifford algebra in $2+1$-dimensional space-time and find that two of them satisfy \eqref{3.4} and
two  other satisfy \eqref{3.5}. However, if one chooses
\begin{equation}\label{3.6}
K=c_1 I+ c_2 \alpha^r
\end{equation}
to satisfy \eqref{3.5}, then \eqref{3.3} is violated. There remains
the only possibility to choose
\begin{equation}\label{3.7}
K= c_1 \gamma^0 + c_2 {\rm i} \gamma^0 \alpha^r
\end{equation}
with real coefficients obeying condition
\begin{equation}\label{3.8}
c_1^2+c_2^2 = 1;
\end{equation}
then both \eqref{3.3} and \eqref{3.4} are satisfied. Using obvious
parametrization $$ c_1 = \sin\theta,\quad c_2= \cos\theta, $$ we
finally obtain
\begin{equation}\label{3.9}
K = {\rm i} \gamma^0 \alpha^r e^{-{\rm i}\theta \alpha^r}.
\end{equation}
Thus, boundary condition \eqref{3.2} with $K$ given by \eqref{3.9}
is the most general boundary condition ensuring self-adjointness
of the Hamiltonian operator on a surface with a deleted disc of radius $r_0$, and parameter $\theta$ can be
interpreted as the self-adjoint extension parameter. Value
$\theta=0$ corresponds to the MIT bag boundary condition which was
proposed as the condition ensuring the confinement of the matter
field, that is, the absence of the matter flux across the boundary
\cite{Joh}. However, it should be comprehended  that a condition
with an arbitrary value of $\theta$ is motivated equally as well as
that with $\theta=0$.

Imposing the boundary condition \eqref{3.2} with matrix $K$
\eqref{3.9} on the solution to the Dirac-Weyl equation,
$\psi_E(\textbf{x})$ \eqref{1.16}, we obtain the condition for the
modes:
\begin{equation}\label{3.10}
\cos\left(\frac{\theta}{2}+\frac{\pi}4 \right)
f_{n,\pm}(r_0,E)=-\sin\left(\frac{\theta}{2}+\frac{\pi}4
\right)g_{n,\pm}(r_0,E).
\end{equation}

Let us consider nanocones with  $N_d=1,\,2,\,3,\,4,\,5$ $\,\,$
($1<\nu<7$), as well as with  $N_d=-1,\,-2,\,-3$ $\,\,$
($\frac{3}{5} < \nu< 1$), and introduce positive quantity
\begin{equation}\label{1.7}
F=\frac32 \nu - \frac12 \nu {\rm sgn} (\nu - 1) - 1,
\end{equation}
which exceeds $1$ at $N_d=3,\,4,\,5$ $\,\,$ ($2\leq\nu<7$) only;
here ${\rm sgn}(u)$ is the sign function, 
${\rm sgn}(u)=1$ at $u>0$ and ${\rm sgn}(u)=-1$ at $u<0$. Define also
\begin{equation}\label{2.0}
n_c=\pm\frac12[{\rm sgn}(\nu-1)-1],
\end{equation}
as well as
\begin{multline}\label{2.5}
\left(
\begin{array}{c}
f_{n_{\rm c}} \\ g_{n_{\rm c}}
\end{array}
\right)= \frac12 \sqrt{\frac{\nu}{\pi} }
\frac1{\sqrt{1+\sin(2\mu_{1-F})\cos(F\pi) } } \\ \times \left(
\begin{array}{c}
 \left[\sin(\mu_{1 -F}) J_{-F}(kr) + \cos(\mu_{1 -F}) J_{F}(kr)\right] \\
{\rm sgn}(E) \left[\sin(\mu_{1 -F}) J_{1-F}(kr) -
\cos(\mu_{1 -F}) J_{-1+F}(kr)\right]
\end{array}
\right),
\end{multline}
\begin{multline}\label{2.1}
\left(
\begin{array}{c}
f_n^{(\wedge)} \\ g_n^{(\wedge )}
\end{array}
\right)= \frac12 \sqrt{\frac{\nu}{\pi} }\\
\times\left(
\begin{array}{c}
 \left[\sin(\mu^{(\wedge)}_{\nu l +1 -F}) J_{\nu l-F}(kr) + \cos(\mu^{(\wedge)}_{\nu l +1 -F}) Y_{\nu l-F}(kr)\right] \\
{\rm sgn}(E)  \left[\sin(\mu^{(\wedge)}_{\nu l +1 -F})
J_{\nu l+1-F}(kr) + \cos(\mu^{(\wedge)}_{\nu l +1 -F}) Y_{\nu
l+1-F}(kr)\right]
\end{array}
\right),
\end{multline}
where $l=n-n_{\rm c}$, and
\begin{multline}\label{2.2}
\left(
\begin{array}{c}
f_n^{(\vee)} \\ g_n^{(\vee )}
\end{array}
\right)=  \frac12 \sqrt{\frac{\nu}{\pi} }\\
\times\left(
\begin{array}{c}
 \left[\sin(\mu^{(\vee)}_{\nu l' +F}) J_{\nu l'+F}(kr) + \cos(\mu^{(\vee)}_{\nu l' +F}) Y_{\nu l'+F}(kr)\right] \\
-{\rm sgn}(E)  \left[\sin(\mu^{(\vee)}_{\nu l' +F})
J_{\nu l'-1+F}(kr) + \cos(\mu^{(\vee)}_{\nu l' +F}) Y_{\nu
l'-1+F}(kr)\right]
\end{array}
\right),
\end{multline}
where $l'=-n+n_{\rm c}$; here $J_\lambda(u)$ and  $Y_\lambda(u)$ are the Bessel
and Neumann functions of order $\lambda$.

In the case of $2 \leq \nu < 7 $ $\quad$ ($F = \nu -1, \,  N_d=3,\,4,\,5$), the complete set of solutions to
\eqref{1.17} is given by
\begin{equation}\label{2.4}
\left. \left( \begin{array}{c} f_{n,\pm} \\ g_{n,\pm}
\end{array}
\right)\right|_{n \geq  n_{\rm c}+1} = \left(
\begin{array}{c}
f_n^{(\wedge)} \\ g_n^{(\wedge )}
\end{array}
\right), \quad \left. \left( \begin{array}{c} f_{n,\pm} \\ g_{n,\pm}
\end{array}
\right)\right|_{n \leq n_{\rm c}} = \left(
\begin{array}{c}
f_n^{(\vee)} \\ g_n^{(\vee )}
\end{array}
\right).
\end{equation}
In the case of $\frac 35 < \nu < 2$ $\quad$ ($0 < F < 1, \,  N_d =2,1,-1, -2, -3$), the complete set of solutions to
\eqref{1.17} is given by
\begin{multline}\label{2.4}
\left. \left( \begin{array}{c} f_{n,\pm} \\ g_{n,\pm}
\end{array}
\right)\right|_{n \geq  n_{\rm c}+1} = \left(
\begin{array}{c}
f_n^{(\wedge)} \\ g_n^{(\wedge )}
\end{array}
\right), \quad \left. \left( \begin{array}{c} f_{n,\pm} \\ g_{n,\pm}
\end{array}
\right)\right|_{n = n_{\rm c}} = \left(
\begin{array}{c}
f_{n_{\rm c}} \\ g_{n_{\rm c}}
\end{array}
\right), \\
\left. \left( \begin{array}{c} f_{n,\pm} \\ g_{n,\pm}
\end{array}
\right)\right|_{n \leq n_{\rm c} - 1} = \left(
\begin{array}{c}
f_n^{(\vee)} \\ g_n^{(\vee )}
\end{array}
\right).
\end{multline}
It should be noted that, in the case of $\nu = \frac12 $ $\quad$ ($N_d=-6$), the complete set of solutions to
\eqref{1.17} is also given by \eqref{2.4} with
$ F = 1/2 $ and $n_{\rm c}=\mp 2 $.

Let us  compare this with the case of an infinitely thin (pointlike) disclination which was considered in detail in
 \cite{SiV7,SiV1,SiV2}. In the latter case several partial Hamiltonian operators are self-adjoint extended,
  and the deficiency index can be $(0,0)$ (no need for extension, all partial operators are essentially self-adjoint),
   $(1,1)$ (one partial operator is extended with one parameter), $(2,2)$  (two partial operators are extended with four parameters),
    etc. In particular, in the case of carbon-like nanocones, there is no need for self-adjoint extension for $N_d=3,4,5$, there is one self-adjoint
    extension parameter for $N_d=2,1,-1$, $-2$, $-3,$ $-6$, there are four and more self-adjoint extension parameters for $N_d=-4,-5$ and $N_d\leq -7$.
    For the deficiency index equal to $(1,1)$, the boundary condition at the location of a pointlike disclination $(r=0)$ takes form
\begin{equation}\label{2.11}
    \lim_{r\rightarrow0}
    \left(\frac{r}{r_{\rm max}}\right)^F \cos\left(\frac{\Theta}{2}+\frac{\pi}{4}\right) f_{{n_{\rm
    c}},\pm} (r,E)=-\lim_{r\rightarrow 0} \left(\frac{r}{r_{\rm max}}\right)^{1-F}\sin\left(\frac{\Theta}{2}+\frac{\pi}{4}\right)
    g_{{n_{\rm
    c}},\pm} (r,E),
\end{equation}
where $\Theta$ is the self-adjoint extension parameter, $F$ is given by \eqref{1.7} for $N_d=2,1,-1$, $-2$, $-3$ and $F=1/2$ for $N_d=-6$, while $n_{\rm c}$ is given by \eqref{2.0} for $N_d=2,1,-1$, $-2$, $-3$ and
$n_{\rm c}=\mp2$ for $N_d=-6$. As follows
from the present section, in the case of a disclination of nonzero size, when the boundary condition is imposed at its edge, the total Hamiltonian operator is self-adjoint extended with the use of one parameter, see \eqref{3.10}.

Value $\Theta$ of the self-adjoint extension parameter in the case of a pointlike disclination can be fixed by the limiting procedure $r_0\rightarrow 0$ in the case of a nonzero-size disclination.
Namely in this way, the condition of minimal irregularity
\cite{Si6,Si7} is obtained:
\begin{equation}\label{6.1}
\Theta=\left\{
\begin{array}{l}
\frac\pi2, \quad 0 < F < \frac12,\\
\vphantom{\int\limits_0^0}
\theta, \quad F = \frac12,\\
-\frac\pi2,\quad \frac12 < F < 1.
\end{array}
\right.
\end{equation}
It should be noted that scale invariance is broken (condition \eqref{2.11} depends on $r_{\rm max}$) unless $\Theta = \pm \pi/2$ at $F \neq 1/2$ and $F = 1/2$ at arbitrary 
$\Theta$. Thus condition \eqref{6.1} is the only one that is consistent with scale invariance.

\section{Induced ground-state effects}

Using the explicit form of modes $f_{n,\pm}$ and $g_{n,\pm}$,
satisfying \eqref{1.17} and \eqref{3.10}, we can calculate the induced ground-state effects of electronic excitations in carbon-like nanocones. Concerning the $R$-current component which is orthogonal to the conical surface, $j^{R}_3$
\eqref{1.1a}, and the electric current angular component, $j_\varphi$
\eqref{1.19a}, they vanish due to the cancellation between modes with $+$ and $-$ subscripts. The calculation of the $R$-current angular component \eqref{1.19} and the  pseudomagnetic field strength \eqref{1.20} in the case of
$\frac35 < \nu < 2$ $\,\,$ ($0<F<1$) and $\nu=\frac12$ $\, \,$  ($ F=1/2$) yields:
\begin{multline}\label{d1}
\left.j^R_\varphi(r)\right|_{F<\frac12,\theta\neq
-\frac\pi2}=-\frac{v}{(2\pi)^2}\frac1r
\left\{
\int\limits_0^\infty \frac{du}{\cosh^2(u/2)} \right.\\  \times
\frac{\sin(F\pi)
\cosh\left[\left(F+\nu-\frac12 \right)u\right]- \sin[(F+\nu)\pi)]
\cosh\left[\left(F-\frac12\right)u\right]}{\cosh(\nu u)-\cos(\nu
\pi)}\\
+8\int\limits_0^\infty dw\,w \left[\sum_{l=0}^\infty
C^{(\wedge)}_{\nu l+1-F}\left(w\frac{r_0}{r}\right)K_{\nu l+1-F}(w)
K_{\nu l-F}(w)\right.\\
-\left.\left.\sum_{l=1}^\infty C^{(\vee)}_{\nu
l+F}\left(w\frac{r_0}{r}\right)K_{\nu l+F}(w) K_{\nu
l-1+F}(w)\right]\right\},
\end{multline}\vspace{-1em}
\begin{multline}\label{d2}
\left.j^R_\varphi(r)\right|_{F>\frac12,\theta\neq
\frac\pi2}=\frac{v}{(2\pi)^2}\frac1r \left\{ \int\limits_0^\infty
\frac{du}{\cosh^2(u/2)} \right.\\  \times \frac{\sin(F\pi)
\cosh\left[\left(F-\nu-\frac12 \right)u\right]- \sin[(F-\nu)\pi)]
\cosh\left[\left(F-\frac12\right)u\right]}{\cosh(\nu u)-\cos(\nu
\pi)}\\
-8\int\limits_0^\infty dw\,w\left[\sum_{l=1}^\infty
C^{(\wedge)}_{\nu l+1-F}\left(w\frac{r_0}{r}\right)K_{\nu l+1-F}(w)
K_{\nu l-F}(w)\right.\\
-\left.\left.\sum_{l=0}^\infty C^{(\vee)}_{\nu
l+F}\left(w\frac{r_0}{r}\right)K_{\nu l+F}(w) K_{\nu
l-1+F}(w)\right]\right\},
\end{multline}\vspace{-1em}
\begin{multline}\label{d3}
\left.j^R_\varphi(r)\right|_{F\neq\frac12,\theta=\pm
\frac\pi2}=\mp\frac{v}{2(2\pi)^2}\frac1r \left\{
\int\limits_0^\infty \frac{du}{\cosh^2(u/2)} \right.\\  \times
\frac{\sin(F\pi) \cosh\left[\left(F-\frac12\pm \nu\right)u\right]-
\sin\left[\left(F \pm
\nu\right)\pi\right]\cosh\left[\left(F-\frac12\right)u\right]}{\cosh(\nu
u)-\cos(\nu
\pi)}\\
+8\int\limits_0^\infty
dw\,w\left[\frac{I_{\frac12\mp\left(F-\frac12\right)}\left(w\frac{r_0}{r}\right)}{K_{\frac12\mp\left(F-\frac12\right)}\left(w\frac{r_0}{r}\right)}
K_{F}(w)
K_{1-F}(w)\right.\\
+\sum_{l=1}^\infty\Biggl( \frac{I_{\nu
l-F+\frac12\pm\frac12}\left(w\frac{r_0}{r}\right)}{K_{\nu
l-F+\frac12\pm\frac12}\left(w\frac{r_0}{r}\right)} K_{\nu l+1-F}(w)
K_{\nu l-F}(w)\Biggr.\\
\left.\left.\Biggl.+\frac{I_{\nu
l+F-\frac12\mp\frac12}\left(w\frac{r_0}{r}\right)}{K_{\nu
l+F-\frac12\mp\frac12}\left(w\frac{r_0}{r}\right)} K_{\nu l+F}(w)
K_{\nu l-1+F}(w)\Biggr)\right]\right\},
\end{multline}\vspace{-1em}
\begin{equation}\label{d4}
\left.j^R_\varphi(r)\right|_{F=\frac12}=-\frac{v\sin\theta}{2\pi^2}\left[\frac1{r-r_0}+\frac8r
\int\limits_0^\infty dw\,w \sum_{l=1}^\infty \tilde C_{\nu
l+\frac12}\left(w\frac{r_0}{r}\right) K_{\nu l+\frac12}(w)K_{\nu
l-\frac12}(w)\right],
\end{equation}\vspace{-1em}
\begin{multline}\label{d5a}
\left.B^{R}_3(r)\right|_{F<\frac12,\theta\neq
-\frac\pi2}=-\frac{\nu}{(2\pi)^2}
\left\{\left(\frac1r
- \frac1r_{\rm max}\right)
\int\limits_0^\infty \frac{du}{\cosh^2(u/2)} \right.\\  \times
\frac{\sin(F\pi)
\cosh\left[\left(F+\nu-\frac12 \right)u\right]- \sin[(F+\nu)\pi)]
\cosh\left[\left(F-\frac12\right)u\right]}{\cosh(\nu u)-\cos(\nu
\pi)}\\
+8\int\limits_r^{r_{max}} \frac{dr'}{r'^2}\int\limits_0^\infty
dw\,w\left[\sum_{l=0}^\infty C^{(\wedge)}_{\nu
l+1-F}\left(w\frac{r_0}{r'}\right)K_{\nu l+1-F}(w)
K_{\nu l-F}(w)\right.\\
-\left.\left.\sum_{l=1}^\infty C^{(\vee)}_{\nu
l+F}\left(w\frac{r_0}{r'}\right)K_{\nu l+F}(w) K_{\nu
l-1+F}(w)\right]\right\},
\end{multline}\vspace{-1em}
\begin{multline}\label{d6}
\left.B^{  R}_3(r)\right|_{F>\frac12,\theta\neq \frac\pi2}=\frac{\nu}{(2\pi)^2} \left\{\left(\frac1r
- \frac1r_{\rm max}\right)
\int\limits_0^\infty \frac{du}{\cosh^2(u/2)} \right.\\  \times
\frac{\sin(F\pi)
\cosh\left[\left(F-\nu-\frac12 \right)u\right]- \sin[(F-\nu)\pi)]
\cosh\left[\left(F-\frac12\right)u\right]}{\cosh(\nu u)-\cos(\nu
\pi)}\\
-8\int\limits_{r}^{r_{max}} \frac{dr'}{r'^2} \int\limits_0^\infty
dw\,w\left[\sum_{l=1}^\infty C^{(\wedge)}_{\nu
l+1-F}\left(w\frac{r_0}{r'}\right)K_{\nu l+1-F}(w)
K_{\nu l-F}(w)\right.\\
-\left.\left.\sum_{l=0}^\infty C^{(\vee)}_{\nu
l+F}\left(w\frac{r_0}{r'}\right)K_{\nu l+F}(w) K_{\nu
l-1+F}(w)\right]\right\},
\end{multline}\vspace{-1em}
\begin{multline}\label{d7}
\left.B^{R}_3(r)\right|_{F\neq\frac12,\theta=\pm
\frac\pi2}=\mp\frac{\nu}{(2\pi)^2}
\left\{\left(\frac1r
- \frac1r_{\rm max}\right)
\int\limits_0^\infty \frac{du}{\cosh^2(u/2)} \right.\\  \times
\frac{\sin(F\pi) \cosh\left[\left(F-\frac12\pm
\nu\right)u\right]- \sin\left[\left(F \pm
\nu\right)\pi\right]\cosh\left[\left(F-\frac12\right)u\right]}{\cosh(\nu u)-\cos(\nu
\pi)}\\
+8\int\limits_{r}^{r_{max}} \frac{dr'}{r'^2} \int\limits_0^\infty
dw\,w\left[\frac{I_{\frac12\mp\left(F-\frac12\right)}\left(w\frac{r_0}{r'}\right)}{K_{\frac12\mp\left(F-\frac12\right)}\left(w\frac{r_0}{r'}\right)}
K_{F}(w)
K_{1-F}(w)\right.\\
+\sum_{l=1}^\infty\Biggl( \frac{I_{\nu
l-F+\frac12\pm\frac12}\left(w\frac{r_0}{r'}\right)}{K_{\nu
l-F+\frac12\pm\frac12}\left(w\frac{r_0}{r'}\right)} K_{\nu l+1-F}(w)
K_{\nu l-F}(w)\Biggr.\\
\left.\Biggl.\left.+\frac{I_{\nu
l+F-\frac12\mp\frac12}\left(w\frac{r_0}{r'}\right)}{K_{\nu
l+F-\frac12\mp\frac12}\left(w\frac{r_0}{r'}\right)} K_{\nu l+F}(w)
K_{\nu l-1+F}(w)\Biggr)\right]\right\}
\end{multline}\vspace{-1em}
and
\begin{multline}\label{d8}
\left.B^{R}_3(r)\right|_{F=\frac12}=\frac{\nu\sin\theta}{2\pi^2}\!\left[\!\frac1{r_0}\ln\left(1\!-\!\frac{r_0}r\right)-\!\frac1{r_0}\ln\left(1\!-\!\frac{r_0}{r_{\rm max}}\right)\right.
\\ \left.-8\!\int\limits_{r}^{r_{max}}\! \frac{dr'}{r'^2}\! \int\limits_0^\infty\! dw\,w
\sum_{l=1}^\infty \tilde C_{\nu
l+\frac12}\left(w\frac{r_0}{r'}\right)\! K_{\nu l+\frac12}(w)K_{\nu
l-\frac12}(w)\right]\!,
\end{multline}\vspace{-1em}
where
\begin{multline}\label{4.6}
C^{(\wedge)}_\rho(y)=\left\{I_\rho(y)K_\rho(y)\tan\left(\frac\theta2+\frac\pi4\right)
-I_{\rho-1}(y)K_{\rho-1}(y)\cot\left(\frac\theta2+\frac\pi4\right)
\right\} \\
\times\left[K^2_\rho(y)\tan\left(\frac\theta2+\frac\pi4\right) + K^2_{\rho-1}(y)\cot\left(\frac\theta2+\frac\pi4\right)
\right]^{-1},
\end{multline}
\begin{multline}\label{4.7}
C^{(\vee)}_\rho(y)=\left\{I_\rho(y)K_\rho(y)\cot\left(\frac\theta2+\frac\pi4\right)-
I_{\rho-1}(y)K_{\rho-1}(y)\tan\left(\frac\theta2+\frac\pi4\right)
\right\} \\
\times\left[K^2_\rho(y)\cot\left(\frac\theta2+\frac\pi4\right) + K^2_{\rho-1}(y)\tan\left(\frac\theta2+\frac\pi4\right)
\right]^{-1}
\end{multline}
and
\begin{equation}\label{4.21}
{\tilde C}_{\nu l+\frac12}(y)=\frac2y\frac{K_{\nu
l+\frac12}(y)K_{\nu l-\frac12}(y)}{\cos^2\theta\left[K_{\nu l+\frac12}^2(y)+K_{\nu l-\frac12}^2(y)\right]^2+4\sin^2\theta \,K^2_{\nu l+\frac12}(y)K^2_{\nu
l-\frac12}(y)};
\end{equation}
$I_\lambda(y)$ and $K_\lambda(y)$ are the modified Bessel functions with
the exponential increase and decrease, respectively, at large real
positive values of their argument.

In the case of $2\leq\nu<7$ $\,\,$ ($F=\nu-1$) we obtain
\begin{multline}\label{d5}
j^R_\varphi(r)=-\frac{v}{(2\pi)^2}\frac1r
\left\{\frac{2\pi}{\nu}\sum_{p=1}^{\left[\!\left| {\nu}/2
\right|\!\right]}
\frac{\sin(3p\pi/\nu)}{\sin^2(p\pi/\nu)}+\frac\pi\nu \delta_{\nu, \,
2N}
+\sin(\nu\pi)\int\limits_0^\infty \frac{du}{\cosh^2(u/2)}\right.\\
\times\frac{ \cosh\left(\frac32 u \right)}{\cosh(\nu u)-\cos(\nu
\pi)} +8\int\limits_0^\infty dw\,w\left[\sum_{l=1}^\infty
C^{(\wedge)}_{\nu (l-1)+2}\left(w\frac{r_0}{r}\right)K_{\nu
(l-1)+2}(w)
K_{\nu (l-1)+1}(w)\right.\\
-\left.\left.\sum_{l=0}^\infty C^{(\vee)}_{\nu
(l+1)-1}\left(w\frac{r_0}{r}\right)K_{\nu (l+1)-1}(w) K_{\nu
(l+1)-2}(w)\right]\right\}
\end{multline}\vspace{-1em}
and
\begin{multline}\label{d6a}
B^{R}_3(r)=-\frac{\nu}{(2\pi)^2}
\left\{\left(\frac1r
- \frac1r_{\rm max}\right)\left[\frac{2\pi}{\nu}\sum_{p=1}^{\left[\!\left| {\nu}/2
\right|\!\right]} \frac{\sin(3p\pi/\nu)}{\sin^2(p\pi/\nu)}+\frac\pi\nu \delta_{\nu, \, 2N} \right. \right.\\
\left. + \sin(\nu\pi)\int\limits_0^\infty \frac{du}{\cosh^2(u/2)}
\frac{ \cosh\left(\frac32 u \right)}{\cosh(\nu u)-\cos(\nu
\pi)} \right]\\
+ 8\int\limits_{r}^{r_{max}} \frac{dr'}{r'^2}
\int\limits_0^\infty dw\,w\left[\sum_{l=1}^\infty C^{(\wedge)}_{\nu
(l-1)+2}\left(w\frac{r_0}{r'}\right)K_{\nu (l-1)+2}(w)
K_{\nu (l-1)+1}(w)\right.\\
\left.\left.-\sum_{l=0}^\infty C^{(\vee)}_{\nu
(l+1)-1}\left(w\frac{r_0}{r'}\right)K_{\nu (l+1)-1}(w) K_{\nu
(l+1)-2}(w)\right]\right\},
\end{multline}
where $\left[\!\left| u \right|\!\right]$ is the integer part of
quantity $u$ (i.e. the integer which is less than or equal to $u$),
$p$ and  $N$ denote positive integers, $\delta_{\omega, \, \omega'}$ is the Kronecker symbol
($\delta_{\omega, \, \omega'}=0$ at $\omega' \neq \omega$ and
$\delta_{\omega, \, \omega} = 1$).

It should be noted that the integral over the $w$ variable in \eqref{d1} -- \eqref{d4} and  \eqref{d5}
vanishes in the limit of $r_0 \rightarrow 0$. Moreover, in the limit of $r \rightarrow \infty$, it decreases
as $(r_0/r)^{2\lambda}$, where
\begin{equation}\label{39}
\lambda=1-F, \quad \frac 35 < \nu <2, \quad \left\{\begin{array}{l}
 0<F<\frac12, \quad \theta\neq -\frac\pi2 \\
 \frac12<F <1, \quad \theta =\frac\pi2
\end{array}\right\},
\end{equation}
\begin{equation}\label{40}
\lambda=F, \quad \frac 35 < \nu <2, \quad \left\{\begin{array}{l}
 \frac12<F<1, \quad \theta\neq \frac\pi2 \\
 0<F <\frac12, \quad \theta =-\frac\pi2
\end{array}\right\},
\end{equation}
\begin{equation}\label{41}
\left\{\begin{array}{l}
\lambda=\nu+\frac 12, \quad \theta\neq \pm \frac\pi2 \\
\lambda=\nu-\frac 12,  \quad \theta= \pm \frac\pi2      \end{array}\right\},
 \quad \frac 35 < \nu <2, \quad F=\frac12,
\end{equation}
\begin{equation}\label{41a}
\left\{\begin{array}{l}
\lambda=1, \hspace{5.7em} \theta\neq \pm \frac\pi2 \\
\lambda=\frac12 \ln \ln (r/r_0),  \quad \theta= \pm \frac\pi2      \end{array}\right\},
 \quad \nu=\frac12, \quad F=\frac12,
\end{equation}
\begin{equation}\label{42}
\left\{\begin{array}{l}
 \lambda=\frac{\ln \ln (r/r_0)}{2\ln (r/r_0)}, \quad \nu=2 \\
 \lambda=\nu-2, \quad 2<\nu<7
\end{array}\right\},\quad F=\nu-1.
\end{equation}
The latter circumstance has far-reaching consequences, when we turn to the total flux of the induced ground-state
pseudomagnetic field strength, see \eqref{1.21}. Namely, the contribution of the $w$-integral to $\Phi_{\rm I}^R$ is
damped and the field strength is proportional  to
the current in the physically sensible case, i.e. at $r_{\rm max} \gg r_0$:
\begin{equation}\label{last1}
j^R_\varphi(r)=\frac{v\Phi_{\rm I}^R}{\pi r_{\rm max}} \,\, \frac1r,
\quad B^{R}_3(r)=\frac{\nu\, \Phi_{\rm I}^R}{\pi r_{\rm max}} \,\,
\left(\frac1r - \frac1r_{\rm max}\right),
\end{equation}
where\vspace{-1em}
\begin{multline}\label{d10}
\left.\Phi_{\rm I}^R \right|_{0<F<\frac12, \,\, \theta\neq
-\frac\pi2}=\left.\Phi_{\rm I}^R\right|_{\frac12<F<1, \,\, \theta=\frac\pi2}=
-\frac{1}{4\pi}\,r_{\rm max} \int\limits_0^\infty \frac{du}{\cosh^2(u/2)} \\  \times
\frac{\sin(F\pi)
\cosh\left[\left(F+\nu-\frac12 \right)u\right]- \sin[(F+\nu)\pi)]
\cosh\left[\left(F-\frac12\right)u\right]}{\cosh(\nu u)-\cos(\nu
\pi)}, \quad \frac 35 < \nu <2,
\end{multline}
\begin{multline}\label{d11}
\left.\Phi_{\rm I}^R\right|_{\frac12<F<1, \,\, \theta\neq
\frac\pi2}=\left.\Phi_{\rm I}^R \right|_{0<F<\frac12, \,\, \theta=-\frac\pi2}=
\frac{1}{4\pi}\,r_{\rm max} \int\limits_0^\infty \frac{du}{\cosh^2(u/2)} \\  \times
\frac{\sin(F\pi)
\cosh\left[\left(F-\nu-\frac12 \right)u\right]- \sin[(F-\nu)\pi)]
\cosh\left[\left(F-\frac12\right)u\right]}{\cosh(\nu u)-\cos(\nu
\pi)}, \quad \frac 35 < \nu <2,
\end{multline}
\begin{equation}\label{d13}
\left.\Phi_{\rm I}^R \right|_{F=\frac12}=-\frac{\sin\theta}{2\pi}r_{\rm max}
\end{equation}
and
\begin{multline}\label{d13}
\left.\Phi_{\rm I}^R \right|_{F=\nu-1}=-\frac{1}{4\pi}r_{\rm max}
\left[\frac{2\pi}{\nu}\sum_{p=1}^{\left[\!\left| {\nu}/2
\right|\!\right]} \frac{\sin(3p\pi/\nu)}{\sin^2(p\pi/\nu)}+\frac\pi\nu \delta_{\nu, \, 2N} \right. \\
\left. + \sin(\nu\pi)\int\limits_0^\infty \frac{du}{\cosh^2(u/2)}
\frac{\cosh\left(\frac32 u \right)}{\cosh(\nu u)-\cos(\nu
\pi)}\right], \quad 2\leq\nu<7.
\end{multline}

The analysis of the induced ground-state electric charge and $P$-condensate is performed in a similar way. Basing on the acquired experience, the results in the physically sensible case ($r_{\rm max} \gg r_0$) can be immediately obtained by employing the condition of minimal irregularity, see \eqref{6.1},
in the case of a pointlike disclination. Note that in the latter case the contribution of modes \eqref{2.1} and \eqref{2.2} is canceled upon summation over the energy sign, thus
\begin{equation}\label{71}
    q(r)=\rho(r)=0, \quad F=\nu-1 \quad (2\leq\nu<7).
\end{equation}
Otherwise, at $0< F<1$ $\quad$ ($\frac35 < \nu <2$ and
$\nu=\frac12$), only mode \eqref{2.5} contributes, and the appropriate results for arbitrary $\Theta$ were
 first obtained in \cite{Si9,Si0} and later generalized to $\nu \neq 1$ in \cite{SiV7,SiV1,SiV2}:\vspace{-0.5em}
\begin{equation}\label{72}
    q(r)=-\frac{e\nu\sin(F\pi)}{\pi^3r^2}\!\int\limits_{0}^{\infty}\! dw\,w \frac{K_F^2(w)-K^2_{1\!-\!F}(w)}
    {{\rm cosh}\!\left[(2F\!-\!1){\rm ln}(w\frac{r_{\rm max}}{r})+{\rm ln}\tan(\frac \Theta 2+\frac \pi 4)\right]}
\end{equation}
and
\begin{equation}\label{73}
    \rho(r)=-\frac{\nu\sin(F\pi)}{\pi^3r^2}\int\limits_{0}^{\infty}dw\,
    w\frac{K_F^2(w)+K_{1-F}^2(w)}{\cosh\left[(2F-1)\ln\left(w\frac{r_{\rm max}}{r}\right)+\ln
    \tan\left(\frac \Theta 2+\frac \pi 4\right)\right]}.
\end{equation}
By appllying  \eqref{6.1} to  \eqref{72} and  \eqref{73} we obtain
\begin{equation}\label{74}
    q(r)=0, \quad 0<F<1,
\end{equation}
\begin{equation}\label{75}
    \rho(r)=0, \quad \left\{\begin{array}{l}
 0<F<\frac12  \\
 \frac12<F <1
\end{array}\right\}
\end{equation}
and
\begin{equation}\label{76}
    \rho(r)=-\frac{\nu\cos \theta}{2\pi^2r^2}, \quad F=\frac12.
\end{equation}
Thus, the electric charge is not induced at all, while the $P$-condensate is induced at $F=1/2$ only, with the total value equal to
\begin{equation}\label{77}
\left. C \right|_{F=1/2}=-\frac{\cos\theta}{\pi} \ln(r_{\rm max}/r_0).
\end{equation}

Recalling the relation between the $P$-condensate and the electric
current, see \eqref{1.19a} and  \eqref{1.19f}, we get that the
induced ground-state electric current density is nonvanishing at
$F=1/2$ only, being directed orthogonally to the conical
surface:\vspace{-0.5em}
\begin{equation}\label{97}
\left. j_3(r) \right|_{F=1/2}=\frac{\nu \left. J_3
\right|_{F=1/2}}{2\pi \ln(r_{\rm max}/r_0)} \, r^{-2},
\end{equation}
where
\begin{equation}\label{98}
\left. J_3 \right|_{F=1/2}=-e v \frac{\cos\theta}{\pi} \ln(r_{\rm max}/r_0)
\end{equation}
is the total electric current. The induced ground-state magnetic field circulating in the angular direction around the apex of the conical surface, see  \eqref{1.20b}, is presented as
\begin{multline}\label{99}
\left. B_\varphi(r) \right|_{F=1/2} - \left. B_\varphi(r_{\rm max}) \right|_{F=1/2} = -\frac{e \left. C
\right|_{F=1/2}}{2\pi \ln(r_{\rm max}/r_0)} \,\ln(r_{\rm max}/r) \\
=-\frac{\left. J_3
\right|_{F=1/2}}{2\pi \ v ln(r_{\rm max}/r_0)} \,\ln(r_{\rm max}/r)=e  \frac{\cos\theta}{2\pi^2} \ln\frac{r_{\rm max}}{r},
\end{multline}
where it is plausible to put the constant of integration equal to zero, $\left. B_\varphi(r_{\rm max}) \right|_{F=1/2} =0$. Then the total magnetic flux, see  \eqref{1.20c}, is
\begin{equation}\label{99}
\left. \Phi_I \right|_{F=1/2}=-e  \frac{\cos\theta}{2\pi^2} r_{\rm max}\left( \ln\frac{r_{\rm max}}{r_0} - 1 + \frac{r_0}{r_{\rm max}}\right).
\end{equation}

\section{Conclusions}

On the basis of the
continuum model for long-wavelength charge carriers, originating in the tight-binding approximation for the nearest-neighbour
interaction of the lattice atoms, we have studied quantum ground-state effects of electronic excitations in crystalline monolayers warped into nanocones by a disclination;
the nonzero size of the disclination at the apex of a nanocone has been taken into account. Our main finding is that the physically sensible limit of the nanocone size exceeding considerably the disclination size fixes a boundary condition at the nanocone apex as the scale invariant one ensuring the minimal irregularity of the modes; consequently, quantum ground-state effects are independent of the disclination size.

Restricting ouselves to the carbon-like nanocones, we have considered all disclinations resulting in the conventional nanocones, $N_d = 1, \, 2, \, 3, \, 4, \, 5$, and several disclinations resulting in the saddle-like nanocones, $N_d = -1, \, -2, \, -3, \, -6$. As we have proved, the results obtained earlier in 
\cite{SiV7,SiV1,SiV2} for the case of a zero-size disclination should be reduced to the case obtained by imposing condition \eqref{6.1}. In particular, the ground-state electric charge is not induced at all. As to the local density of states, it is defined as 
\begin{equation}\label{1.121}
\Delta(\textbf{x};E')= \int\limits_{-\infty}^\infty \frac{dE\,
|E|}{\pi\hbar^2 v^2} \, \mbox{$\psi$}^\dag _E(\textbf{x}) \, {\rm Im}(E-E'- {\rm i} 0)^{-1} \, \mbox{$\psi$}
_E(\textbf{x}). 
\end{equation}
The density of the induced ground-state electric charge is related to \eqref{1.121} as 
\begin{equation}\label{1.122}
q(\textbf{x})=-\frac{e}2 \int\limits_{-\infty}^\infty dE'\,
\Delta(\textbf{x};E') \,      
{\rm sgn}(E'),
\end{equation}
and only the odd in $E'$ piece of $\Delta(\textbf{x};E')$ contributes to $q(\textbf{x})$. In the case of planar crystalline monolayer $(\nu=1)$, one immediately gets
\begin{equation}\label{1.123}
\Delta(\textbf{x},E') = \frac{|E'|}{\pi\hbar^2 v^2},
\end{equation}
and, as follows from the nullification  of the charge, disclinations leave relation \eqref{1.123} unchanged; this also follows from expression (55) in \cite{SiV7} for the total density of states (when condition \eqref{6.1} is imposed).

As to the nonvanishing ground-state effects which are induced in carbon-like nanocones, they comprise two sets. One includes the magnetic field circulating in the angular direction around the nanocone apex, the electric current directed orthogonally to the nanocone surface and the parity-breaking condensate. In terms of the sublattice and valley indices, this set corresponds to bilinear form 
$$ \left(
\begin{array}{c}
(I)  \\
A
\end{array}
\right) \,\left(
\begin{array}{c}
(II)  \\
B
\end{array}
\right)\, + \,  \left(
\begin{array}{c}
(I)  \\
B
\end{array}
\right) \,\left(
\begin{array}{c}
(II)  \\
A
\end{array}
\right)$$
and emerges at $F=1/2$ only, i.e. at $N_d = \pm 2, \, -6$. Another set includes the pseudomagnetic field directed orthogonally to the nanocone surface and the $R$-current circulating in the angular direction around the nanocone apex. In terms of the sublattice and valley indices, this set corresponds to bilinear form 
$$ \left(
\begin{array}{c}
(I)  \\
A
\end{array}
\right) \,\left(
\begin{array}{c}
(II)  \\
A
\end{array}
\right)\, - \,  \left(
\begin{array}{c}
(I)  \\
B
\end{array}
\right) \,\left(
\begin{array}{c}
(II)  \\
B
\end{array}
\right)$$
and emerges in all considered cases except $\nu = 3$, i.e. $N_d = 4$. We summarize our results by presenting expressions for the total magnetic and pseudomagnetic fluxes, 
$\Phi_{\rm I}$ and $\Phi_{\rm I}^R$: 
\begin{equation}\label{last2}
\left.\Phi_{\rm I}^R \right|_{\theta\neq -\frac\pi2}\!=\!
-\frac{1}{4\pi}\,r_{\rm max}\!\! \int\limits_0^\infty\!\!
\frac{du}{\cosh^2(u/2)} \, \frac{\sin\left(\frac15\pi\right)
\cosh\left(\frac9{10} u\right)\!-\! \sin\left(\frac75 \pi\right)
\cosh\left(\frac3{10}u\right)}{\cosh\left(\frac65
u\right)-\cos\left(\frac65 \pi\right)},\,\, N_d=1,
\end{equation}\vspace{-1em}
\begin{equation}\label{last3}
\left.\Phi_{\rm I}^R \right|_{\theta= -\frac\pi2}=
\frac{1}{4\pi}\,r_{\rm max} \sin\left(\frac15\pi\right)
\int\limits_0^\infty \frac{du}{\cosh^2(u/2)}\,
\frac{\cosh\left(\frac3{2} u\right)}{\cosh\left(\frac65
u\right)-\cos\left(\frac65 \pi\right)},\,\, N_d=1,
\end{equation}\vspace{-1em}
\begin{equation}\label{last4}
\left.\Phi_{\rm I}^R \right|_{\theta\neq \frac\pi2}\!=\!
\frac{1}{4\pi}\,r_{\rm max}\!\! \int\limits_0^\infty\!\!
\frac{du}{\cosh^2(u/2)} \, \frac{\sin\left(\frac57 \pi\right)
\cosh\left(\frac9{14} u\right)\!+\! \sin\left(\frac17 \pi\right)
\cosh\left(\frac3{14}u\right)}{\cosh\left(\frac67
u\right)-\cos\left(\frac67 \pi\right)},\,\, N_d=-1,
\end{equation}\vspace{-0.5em}
\begin{equation}\label{last5}
\left.\Phi_{\rm I}^R \right|_{\theta= \frac\pi2}\!=\!
-\frac{1}{4\pi}\,r_{\rm max}\!\! \int\limits_0^\infty\!\!
\frac{du}{\cosh^2(u/2)} \frac{\sin\left(\frac57\pi\right)
\cosh\left(\frac{15}{14} u\right)\!-\! \sin\left(\frac{11}7
\pi\right) \cosh\left(\frac3{14}u\right)}{\cosh\left(\frac67
u\right)-\cos\left(\frac67 \pi\right)},\,\, N_d=-1,
\end{equation}\vspace{-0.5em}
\begin{equation}\label{last6}
\Phi_I =- e  \frac{\cos\theta}{2\pi^2} r_{\rm max}\left( \ln\frac{r_{\rm max}}{r_0} - 1 \right), \quad \Phi_{\rm I}^R =-\frac{\sin\theta}{2\pi} \,  r_{\rm max}, \quad  N_d=\pm 2,-6,
\end{equation}\vspace{-0.5em}
\begin{equation}\label{last7}
\left.\Phi_{\rm I}^R \right|_{\theta\neq -\frac\pi2}= -\frac{\sqrt{3}}{8\pi}\,r_{\rm max} \int\limits_0^\infty \frac{du}{\cosh(u/2)}
\frac{1}{\cosh\left(\frac23 u\right)-\cos\left(\frac23
\pi\right)}, \quad  N_d=-3,
\end{equation}\vspace{-0.5em}
\begin{equation}\label{last8}
\left.\Phi_{\rm I}^R \right|_{\theta= -\frac\pi2}= \frac{\sqrt{3}}{8\pi}\,r_{\rm max} \int\limits_0^\infty
\frac{du}{\cosh^2(u/2)} \frac{ \cosh\left(\frac{5}{6} u\right)+
\cosh\left(\frac1{6}u\right)}{\cosh\left(\frac23
u\right)-\cos\left(\frac23 \pi\right)}, \quad  N_d=-3,
\end{equation}\vspace{-0.5em}
\begin{equation}\label{last9}
  \Phi_{\rm I}^R  =\frac{1}{8} \, r_{\rm max}, \quad  N_d=3,
\end{equation}
\begin{equation}\label{last10}
 \Phi_{\rm I}^R  = -\frac{7}{24} \, r_{\rm max}, \quad   N_d=5.
\end{equation}

We conclude that the induced ground-state effects change drastically as $N_d$ changes. The effects are absent in the case of
the four-heptagonal defect ($N_d= 4$), whereas they appear of opposite signs as a heptagon is removed from ($N_d= 3$)
or added to ($N_d= 5$) this defect, see \eqref{last9} and \eqref{last10}. These cases are independent of the boundary parameter, $\theta$; note that namely these cases
correspond to that situation with the zero-size defect when there is no need for self-adjoint extension (the deficiency index is
(0,0)). In all other cases the results depend on $\theta$. The most distinct dependence is characteristic for the cases of
two-pentagonal, two- and six-heptagonal defects, when the results coincide, see \eqref{last6}. In the cases of one-pentagonal, one- and three-heptagonal defects, the results are almost
independent of $\theta$ unless $\theta=-\frac{\pi}{2}$ for $N_d=1, -3$ and $\theta=\frac{\pi}{2}$ for $N_d=-1$, see
\eqref{last2} -- \eqref{last5}, \eqref{last7} and \eqref{last8}.

Effective magnetic and pseudomagnetic fields wchich appear in corrugated crystalline monolayers produce strains and scattering of electronic excitations in a sample \cite{Voz}. As follows from our
consideration, the ground-state magnetic and pseudomagnetic fields can be induced in the locally flat regions out of disclinations, and this may have
observable consequences in experimental measurements, likely with the use of scanning tunnel and transmission electron microscopy.

\phantom{jvhvj}

{\it Acknowledgements}
The work of Yu.A.S. was supported by the National Academy of Sciences of Ukraine (Project No.01172U000237),
by the Program of Fundamental Research of the Department of Physics and Astronomy of the National Academy of
Sciences of Ukraine (Project No.0117U000240) and by the ICTP -- SEENET-MTP project NT-03
`Cosmology - Classical and Quantum Challenges'.



\end{document}